\documentclass[10pt]{article}

\usepackage{amsmath}
\usepackage{amssymb}
\usepackage{graphics}
\usepackage{rotating}
\usepackage{cite}
\usepackage{color}
\usepackage{fancybox}

\def\0{\mbox{\tiny $0$}}
\def\1{\mbox{\tiny $1$}}
\def\2{\mbox{\tiny $2$}}
\def\3{\mbox{\tiny $3$}}
\def\4{\mbox{\tiny $4$}}
\def\5{\mbox{\tiny $5$}}
\def\6{\mbox{\tiny $6$}}
\def\7{\mbox{\tiny $7$}}
\def\8{\mbox{\tiny $8$}}
\def\9{\mbox{\tiny $9$}}
\def\M{\mbox{\tiny $M$}}
\def\Spm{\mbox{\tiny \sc Spm}}
\def\New{\mbox{\tiny \sc New}}

\def\T{\mbox{\tiny $T$}}

\def\tra{\mbox{\tiny $tra$}}

\def\mi{\mbox{\tiny $-$}}

\title{ \shadowbox{\large \bf A NEW PHASE TIME FORMULA FOR OPAQUE
BARRIER TUNNELING}}
\author{
\small  Stefano De Leo\thanks{Department of Applied Mathematics,
State University of Campinas, Brazil [deleo@ime.unicamp.br] } \,\,
and\, Vinicius Leonardi\thanks{Department of Applied Mathematics,
State University of Campinas, Brazil}}

\date{\small
\fcolorbox{black}{yellow} {\color{red} $\bullet$ {\color{black}{
{\footnotesize  {\sc Journal of Physics A: Math. Theor.} {\bf 44}
(2011) 085305-9}}} {\color{red}{$\bullet$}} } }

\begin{document}
%
%%%%%%%%%%%%%%%%%%%%%%%%%%%%%%%% PAPER %%%%%%%%%%%%%%%%%%%%%%%%%%%%%%%%%%%%%

\maketitle

\vspace*{-.7cm}

\begin{abstract}
\noindent After a brief review of the derivation of the standard
phase time formula, based on the use of the stationary phase
method, we propose, in the opaque limit, an alternative method to
calculate the phase time.  The {\em new} formula for the phase
time  is in excellent agreement with the numerical simulations and
shows that for wave packets whose upper limit of the momentum
distribution is very close to the barrier height, the transit time
is {\em proportional} to the barrier width.
\end{abstract}

%%%%%%%%%%%%%%%%%%%%%%%%%%%%%%%%%%%%%%%%%%%%%%%%%%%%%%%%%%%%%%%%%%%%%%
%%%%%%%%%%%%%%%%%%%%%%%%%%%%%%%%%%%%%%%%%%%%%%%%%%%%%%%%%%%%%%%%%%%%%%%

% Warning: No PACS code given

%\small {\sc Pacs} 03.65.-w, 03.65.Xp, 03.65.Ta.\\

%02.10.Hh Rings and algebras
%02.10.Ud Linear algebra
%02.10.Yn Matrix theory

%02.30.Hq Ordinary differential equations
%02.30.Jr Partial differential equations
%02.30.Tb Operator theory

%03.65.-w Quantum mechanics
%03.65.Ca Formalism
%03.65.Ta Foundations of quantum mechanics;
%03.65.Xp Tunnelling, traversal time, quantum Zeno dynamics

%12.15.F Quarks and lepton masses and mixing
%14.60.Pq Neutrino mass and mixing

%\offprints{~Stefano De Leo.}

%\titlerunning{\sc  opaque barrier tunneling times}

%%%%%%%%%%%%%%%%%%%%%%%%%%%%%%%%%%%%%%%%%%%%%%%%%%%%%%%%%%%%%%%%%%%%%%%
%%%%%%%%%%%%%%%%%%%%%%%%%%%%%%  SECTION   %%%%%%%%%%%%%%%%%%%%%%%%%%%%%
%%%%%%%%%%%%%%%%%%%%%%%%%%%%%%%%%%%%%%%%%%%%%%%%%%%%%%%%%%%%%%%%%%%%%%

\section*{\normalsize I. INTRODUCTION}
The time spent by a particle to tunnel across a barrier surely
represents one of the most intriguing and challenging discussions
found in literature. After the stimulating articles of
MacColl\cite{MC32} and
 Hartman\cite{HAR62} on the dynamics of the  wave packet which
tunnel  potential barriers, many tunneling time definitions have
been introduced and paradoxical effects, such as superluminal
velocities, discussed. Of special interest for us is the
discussion on the time spent by non-relativistic particles to
cross the classical forbidden region. The extensive literature on
tunneling times is reviewed in many reports. For a detailed
discussion on phase, dwell  and Larmor times we refer the reader
to the report of Hauge and Stovneng\cite{HS89}, for photon and
particle tunneling reviewed by a unified time analysis to the
report of Olkhovsky, Recami and Jakiel\cite{REC04},  and, finally,
for a clear, comprehensive  and complete discussion of the
tunneling time definitions, paradoxes and proposed solutions to
the excellent work of Winful\cite{WIN06}.  In this report, it is
also found a challenging elctromagnetic analogy  with the
frustrated total internal reflection and resonant tunneling.

In this paper, we present a detailed analytic and numerical
analysis of the phase time for non relativistic wave packets which
tunnel opaque barriers.
 In the opaque limit, due to the filter effect, approximations on the transmission
coefficient allow  to find a closed formula for the time in which
the peak of the transmitted wave appears in the free region after
the barrier. The {\em new} formula, which generalizes the well
known formula obtained by the stationary phase method, shows that,
for momentum distributions whose upper limit is very close to the
barrier height, the phase time
 is {\em proportional} to the barrier width. The study is done for a potential barrier
 $V_{\0}$ with discontinuity in
$x=0$ and $x=L$.

The method commonly used in calculating the  transmitted amplitude is based on
solving, separately, the Schr\"odinger equation for stationary states within
the potential region and  in the free regions before and after the barrier
and, then, imposing the continuity of the wave function and its derivative at
the discontinuities of the potential. After simple algebraic computations, we
find the following transmitted amplitude\cite{WIN06,MER}
\begin{equation}
T(k) =e^{-ikL}\,\mbox{\huge$/$}\,\left[\,\cosh (qL) - i\,
\frac{2\,k^{^{2}}-w^{\,\2}}{2\,k\,q}\,\sinh (qL) \,\right]\,\,,
\end{equation}
where $q=\sqrt{w^{^{2}}-k^{^{2}}}$ and $w=\sqrt{2\,mV_{\0}}/\hbar$. The
resultant transmitted wave packet is obtained by integrating over all the
possible stationary states modulated by a weighting function $g(k)$,
\begin{equation}
\label{traw} \Phi_{\T}(x,t) =
N\,\int_{\0}^{k_{\M}}\,\hspace*{-.4cm}\mbox{d}k\,g(k)\,\,|T(k)|\,\,e^{i\,\varphi}\,
\,e^{i\,[k(x-L)-Et/\hbar]}\,\,,
\end{equation}
where $\varphi-kL$, with  $\varphi=\arctan[\,(2\,k^{^{2}}-w^{\2})
\tanh (qL)\,/\,2\,k\,q\,]$, is the phase of the transmitted
amplitude,   $k_{\M}=\sqrt{2\,mE_{\M}}/\hbar$ is the upper limit
of the momentum distribution, and $N$ is a normalization constant
containing the information of the number of incoming electrons.
The condition $k_{\M}\leq w$ guarantees that the allowed energies
are truncated before or at the barrier height $V_{\0}$. We do not
have  above barrier contributions and, consequently, only
 tunneling is responsible for the transmitted wave.
We restrict the discussion to pure tunneling, i.e. $E_{\M}\leq V_{\0}$, to
avoid the phenomenon of multiple diffusion\cite{MD1,MD2}.

 As  is  well known\cite{REC04,WIN06},  the use of
 the stationary phase method allows to calculate the
 time in which the peak of the transmitted wave appears in the
 free region after the barrier
 without explicitly solving the integral of Eq.(\ref{traw}). Unfortunately,
this is not sufficient for determining the transit time. In fact,
the use of the standard phase time formula requires a careful
analysis on the applicability of the stationary phase method.
Without such an analysis its indiscriminate use could result in
wrong theoretical  interpretations on the dynamics of the
particles tunneling.

 In the next section, we briefly revise the standard derivation
of the phase time based on the use of the stationary phase method. Then, in
section III, we obtain, by an analytic study of the transmitted wave,  a {\em
new} formula for the phase time . This is done in the opaque limit. The
discrepancy between the standard and the new formula for the phase time is
clear for $E_{\M}$ close to $V_{\0}$. To confirm the validity of the new
formula, a comparison between analytical results and numerical data is
presented in section IV. The agreement is excellent and suggests the use of
the new approach proposed in this paper as a new method for estimating the
phase time. Our conclusions and possible future investigations are drawn in
the final section.

\section*{\normalsize II. REVISING THE STANDARD PHASE TIME FORMULA}

The old question of tunneling times is often addressed by studying
the phase time through the use of the stationary phase
method\cite{JJ,AEI}.
 This method provides an approximate way to
calculate the maximum of an integral. The main idea  is that
sinusoids with rapidly changing phases will add destructively.
This basic principle of asymptotic analysis allows to find the
maximum of an integral independent of the details integrand shape.
Such a maximum depends on the derivative of the integrand {\em
phase} calculated at the mean value of the wave number. For the
transmitted wave given in Eq.(\ref{traw}), the phase term is {\em
stationary} for
\begin{equation}
\label{spm} \left[\,\varphi+ k\,(x-L) -\frac{E\,t}{\hbar}\,
\right]^{\prime}_{\langle k \rangle_{_{T}}} = 0\,\,,
\end{equation}
where the prime stands for the derivative with respect to $k$ and
\begin{equation}
\label{kt} \langle k \rangle_{\T}=
\int_{\0}^{k_{\M}}\,\hspace*{-.2cm}\mbox{d}k\,k\,g^{\2}(k)\,|T(k)|^{^{2}}\mbox{\huge
/}\hspace*{-.1cm}
\int_{\0}^{k_{\M}}\,\hspace*{-.2cm}\mbox{d}k\,g^{\2}(k)\,|T(k)|^{^{2}}\,\,.
\end{equation}
By using Eq.(\ref{spm}), we find that the phase of the transmitted
wave is stationary at $x=L$ for times which satisfy
\begin{equation}
\label{spm2} \left\{\,\frac{E\,t}{\hbar} -
\frac{k\,[w^{\4}\sinh(2\,qL)+2\,k^{\2}\,(w^{\2}-2\,k^{\2})\,qL]}{q
\,[w^{\4}\cosh(2\,qL)] +
8\,k^{\2}q^{\2}-w^{\4}]}\,\right\}_{\langle k
\rangle_{_{T}}}=\,\,0\,\,.
\end{equation}

For thin barriers ($k_{\M}L\ll 1$) the modulus of the transmitted
coefficient is close to $1$. The transmitted wave packet has the
same form of the incident packet and, consequently,  $\langle k
\rangle_{\T}\approx k_{\0}$, where $k_{\0}$ is the center of the
incoming momentum distribution $g(k)$. The phase time is then
proportional to $L$. Observe that for very thin barriers we can
always guarantee $q_{\0}L\ll 1$.

For much thicker barriers ($k_{\M}L\gg 1$), we enter in the
so-called {\em opaque limit}. In this limit, the peak of the
transmitted momentum distribution  is shifted to higher wave
numbers. This effect is known in literature as filter
effect\cite{HAR62}. Before beginning our discussion on the phase
time formula (\ref{spm2}), let us briefly discuss the filter
effect. In the opaque limit, the modulus of the transmitted
coefficient can be approximated by
\begin{equation}
\label{apT} |\,T(k)\,|\,\,\approx\,\,
4\,k\,q\,e^{-q\,L}\,/\,w^{^{2}}\,\,.
\end{equation}
By using this approximation, which also implies $g(k)|T| \approx
g(k_{\M})|T|$, and changing in Eq.\,(\ref{kt}) the variable of integration
from $k$ to $q$, we obtain
\begin{equation}
\label{iap} \langle k \rangle_{\T} \approx
\int_{q_{\M}}^{w}\,\hspace*{-.2cm}\mbox{d}q\,k^{\2}\,q^{\3}\,
e^{-2qL}\mbox{\huge
/}\hspace*{-.1cm}
\int_{q_{\M}}^{w}\,\hspace*{-.2cm}\mbox{d}q\,k\,q^{\3}\,e^{-2qL}\,\,,
\end{equation}
where $q_{\M}=\sqrt{w^{^{2}}-k_{\M}^{^{2}}}$. The integrands can
be now expanded in series around $q_{\M}$,
\begin{eqnarray*}
k^{\2} q^{\3} & = & k_{\M}^{\2}q_{\M}^{\3} + (3\,w^{\2} -
5\,q_{\M}^{\2}) \,q_{\M}^{\2}\,(q-q_{\M})\,+\,
\mbox{O}\left[\left(q-q_{\M}\right)^{^{2}}\right]\,\,,\\
k\,q^{\3} & = & k_{\M}\,q_{\M}^{\3} + (3\,w^{\2} - 4\,q_{\M}^{\2})
\,q_{\M}^{\2}\,(q-q_{\M})\,/\,k_{\M} \,+\,
\mbox{O}\left[\left(q-q_{\M}\right)^{^{2}}\right]\,\,.
\end{eqnarray*}
Observing that the main contribution to the integrals in
Eq.(\ref{iap}) comes from the lower limit $q_{\M}$, we obtain
\begin{equation}
 \langle k \rangle_{\T} \approx
\frac{k_{\M}^{\2}q_{\M} + (3\,w^{\2} - 5\,q_{\M}^{\2})
\,/\,2L}{k_{\M}\,q_{\M} + (3\,w^{\2} - 4\,q_{\M}^{\2})
\,/\,2k_{\M}L} \approx k_{\M} - \frac{q_{\M}}{2\,k_{\M}L}\,\,.
\end{equation}
%When the upper limit of the incoming momentum distribution
%coincides with the barrier wave number, i.e. $k_{\M}=w$, we have
%$q_{\M}=0$. Consequently, the series expansions must be
%re-calculated around $q_{\M}=0$,
%\begin{eqnarray*}
%k^{\2} q^{\3} & = & w^{\2} q^{\3}  - q^{\5}\,\,,\\
%k\,q^{\3} & = & w\,q^{\3} - q^{\5}/\,2\,w\,+\,
%\mbox{O}\left[\,q^{\7}\,\right]\,\,.
%\end{eqnarray*}
%In this case, the mean value of the transmitted momentum becomes
%\begin{equation}
%\langle k \rangle_{\T} \approx \frac{3\,w^{^{2}}- 5\,/\,L^{^{2}}}{
%3\,w- 5\,/\,2wL^{^{2}}}\approx w - \frac{5}{2\,wL^{^{2}}}\,\,.
%\end{equation}
The phase of the transmitted wave is then stationary at $x=L$ for
times which satisfy
\begin{equation}
\label{ptf} \frac{E_{\M}\,t_{\Spm}}{\hbar}\,\approx\,
\frac{k_{\M}}{q_{\M}} \,\,.
\end{equation}
%This implies superluminal transit velocities. Of special interest
%is the limit $k_{\M}\to w$. In this case, we find
%\begin{equation}
%\label{ptf2} \frac{V_{\0}\,t_{\Spm}}{\hbar}\,\approx\,
%\frac{w\,L}{\sqrt{5}} \,\,.
%\end{equation}
%The phase time is now {\em proportional} to $L$ and the transit
%velocity tends to a constant value.
In the next section, we shall propose a new method to calculate
the phase time. The {\em new} phase time formula which reproduces
Eq.(\ref{ptf}) for $q_{\M}\sim k_{\M}$ foresees transit times
which are proportional to the barrier width $L$  for $q_{\M}\ll
k_{\M}$.

\section*{\normalsize III. PROPOSING A  NEW PHASE TIME FORMULA}

In the previous section, we have estimated the position of the
maximum of the transmitted wave packet by using the stationary
phase method. In this section, we propose an {\em alternative}
method to calculate the phase time formula. For opaque barrier
tunneling it is possible to calculate explicitly the derivative
with respect to time of the electronic density at the barrier edge
and finding when it is equal to zero. This allows us to obtain a
{\em new} formula for the time in which the maximum of the
transmitted wave packet is found at $x=L$.

For opaque barrier tunneling, in solving the integral which
appears in Eq.(\ref{traw}) we can use  the approximation given in
Eq.(\ref{apT}) and change the variable of integration from $k$ to
$q$. Consequently, the expression of the transmitted wave at the
edge of the barrier ($x=L$) becomes
\begin{equation}
 \Phi_{\T}(L,t) \approx
4\,N\,g(k_{\M})\,\int_{q_{\M}}^{w}\,\hspace*{-.3cm}\mbox{d}q\,q^{\2}\,e^{-qL}\,
\,e^{i\,(\varphi-Et/\hbar)}\,/\,w^{\2}\,\,.
\end{equation}
Due to the filter effect, the phase $\varphi$ and the energy $E$
can be expanded as follows
\begin{eqnarray*}
\varphi & =& \varphi_{\M} - 2\,(q-q_{\M})\,/\,k_{\M} -
q_{\M}\,(q-q_{\M})^{^2}\,/\,k_{\M}^{^{3}}+\mbox{O}\left[\left(q-q_{\M}\right)^{^{3}}
\right]\,\,,\\
E & = & E_{\M} - \hbar^{\2}\,q_{\M}\,(q-q_{\M})\,/\,m
-\hbar^{\2}\, (q-q_{\M})^{^{2}}\,/\, 2\,m\,\,.
\end{eqnarray*}
Introducing the new adimensional variable
$\rho(q)=(q-q_{\M})/k_{\M}$ and  using the previous expansions, we
obtain for the electronic density at $x=L$,
\[
|\Phi_{\T}(L,t)|^{^{2}}\,\,\approx\,\,16\,|N|^{^{2}}\,
k^{^{6}}_{\M}\,g^{\2}(k_{\M})\,\,\underbrace{\,\left|\,\int_{\0}^{\rho(w)}\,
\hspace*{-.2cm}\mbox{d}\rho\,
\left(\rho+\frac{q_{\M}}{k_{\M}}\right)^{^{2}}\,\exp\{-\rho
\,k_{\M}L
+i\,\,[\,\,\alpha(t)\,\rho\,+\,\beta(t)\rho^{\2}\,\,]\,\,\}\,\right|^{^{2}}}_{\mbox{
\normalsize{$S(t)$}}}\,/\,w^{\4}\,\,,
\]
with
\begin{equation}
\alpha(t) =2\,\left(\, \frac{q_{\M}E_{\M}\,t}{\hbar\, k_{\M}} -
1\right)\,\,\,\,\,\mbox{and}\,\,\,\,\,\beta(t) =
\frac{E_{\M}\,t}{\hbar} - \frac{q_{\M}}{k_{\M}}\,\,.
\end{equation}
The subject matter of this section will be the accurate analysis
of $S(t)$ and the calculation of its derivative with respect to
time. A first approximation is to consider the first terms in the
expansion of the phase time exponential,
\begin{equation}
S(t)\approx \left|\,s(0) + i\, \left[\,\alpha(t)\,s(1)+
\beta(t)\,s(2)\,\right] -
\mbox{$\frac{1}{2}$}\,\left[\,\alpha^{\2}(t)\,s(2) +
2\,\alpha(t)\beta(t)\,s(3) +
\beta^{\2}(t)s(4)\,\right]\,\right|^{^{2}}\,\,,
\end{equation}
where
\[
s(n)=\int_{\0}^{\widetilde{w}}\,\hspace*{-.2cm}\mbox{d}\rho\,
\left(\rho+\frac{q_{\M}}{k_{\M}}\right)^{^{2}}\rho^{n}\,\exp[-\rho\,
k_{\M}L] \approx\,\,
\frac{(n+2)!}{\,\,\,\,\,(k_{\M}L)^{^{n+3}}}\,\left[\,1+2\,
\frac{q_{\M}L}{n+2} +
\frac{(q_{\M}L)^{^{2}}}{(n+2)(n+1)}\right]\,\,.
\]

\noindent $\bullet$ \fbox{\fbox{The case $q_{\M}\sim k_{\M}$}}\\

\noindent
 If we limit ourselves to
the analysis of processes in which $q_{\M}$ is of the order of
$k_{\M}$, we find that $\beta(t)$ is of the same order of
$\alpha(t)$. Observing that in the opaque limit $s(n)\gg s(n+1)$
and that the $\beta$-term in $S(t)$ is coupled to $s(n)$ with
higher $n$, we can approximate $S(t)$ as follows
\begin{equation*}
S(t;q_{\M}\sim k_{\M})  \approx \left|\,s(0) + i\, \alpha(t)\,s(1)
- \mbox{$\frac{1}{2}$}\,\alpha^{\2}(t)\,s(2) \,\right|^{^{2}}
\approx  \,\,s^{\2}(0) +
\alpha^{\2}(t)\,\left[\,s^{\2}(1)-s(0)s(2)\,\right]\,\,.
\end{equation*}
Deriving $S(t;q_{\M}\sim k_{\M})$ with respect to time and
equating to zero, we obtain $\alpha(t)=0$. Thus, in this limit, we
reproduce the well known stationary phase condition (\ref{ptf}),
\begin{equation}
\label{s1} S_t(t;q_{\M}\sim k_{\M})=0\,\,\,\,\,\Rightarrow
\,\,\,\,\,\alpha(t)=0\,\,\,\,\,\Rightarrow
\,\,\,\,\,\frac{E_{\M}t}{\hbar} = \frac{k_{\M}}{q_{\M}}\,\,.
\end{equation}

\noindent $\bullet$ \fbox{\fbox{The case $q_{\M}\ll k_{\M}$}}\\

\noindent In this limit, $\alpha\approx - 2$ and $\beta(t)\approx
V_{\0}t/\hbar$. The $\beta$-term cannot be neglected because for
time of the order of $\hbar L/E_{\M}$ it  becomes comparable to
the $\alpha$-term. This implies in our approximation that the
terms $s(n)$, $t\,s(n+1)$, and $t{^{2}}\,s(n+2)$ are of the same
order. Consequently,
\begin{eqnarray*}
S(t;q_{\M}\ll k_{\M})
     &\approx  &  \left\{\,  s(0) -
\frac{1}{2}\,\left[\,4\,s(2)-4\,s(3)\,\frac{V_{\0}t}{\hbar} +
s(4)\,
\frac{V^{^{2}}_{\0}t^{^{2}}}{\hbar^{^{2}}}\,\right]\right\}^{^{2}}+
\,\, \left[\,-2\,s(1)+
\frac{V_{\0}t}{\hbar}\, \,s(2)\,\right]^{^{2}} \\
 & \approx & s^{\2}(0) + \,4\,
[\,s^{\2}(1)-s(0)s(2)\,]+4
\,[\,s(0)s(3)-s(1)s(2)\,]\,\frac{V_{\0}t}{\hbar}\,+\,
[\,s^{\2}(2)-s(0)s(4)\,]\,\frac{V^{^{2}}_{\0}t^{^{2}}}{\hbar^{^{2}}}\,\,.
\end{eqnarray*}
By taking the derivative with respect to time and setting it equal
to zero, we obtain
\begin{equation}
\label{s2} S_t(t;q_{\M}\ll
k_{\M})=0\,\,\,\,\,\Rightarrow\,\,\,\,\, \frac{V_{\0}t}{\hbar}=
\frac{ 2\,[s(1)s(2)-s(0)s(3)]}{ [s^{\2}(2)-s(0)s(4)]}\approx
\frac{2\,w L}{9} \,\,.
\end{equation}
The transit velocity, defined as the ratio between the barrier
width and  the time in which the peak appears in the free region
after the barrier, is then given by
\begin{equation}
\label{vtra} v_{\mbox{\tiny tra}}=
\frac{9}{2}\,\sqrt{\frac{V_{\0}}{2\,m}}\,\,.
\end{equation}
This analytic result is confirmed by numerical calculations, see
Table 1. The details of our numerical simulations are found in
section IV.\\

\noindent $\bullet$ \fbox{\fbox{The general case}}\\

\noindent Observing  that for increasing times the terms
$\alpha(t)s(n)$, $\beta(t)\,s(n+1)$, and $\beta^{^{2}}(t)\,s(n+2)$
become comparable, we obtain for $S(t)$ the following expression
\begin{equation} S(t)  \approx
s^{\2}(0) + \alpha^{\2}(t) \,[\,s^{\2}(1)-s(0)s(2)\,]+\,
2\,\alpha(t)\,\beta(t)\,[\,s(1)s(2)-s(0)s(3)\,]+ \beta^{^{2}}(t)\,
[\,s^{\2}(2)-s(0)s(4)\,]\,\,.
\end{equation}
In deriving $S(t)$, we use $\alpha_t(t)=2\,q_{\M}E_{\M}/\hbar
k_{\M}$ and $\beta_t(t)=E_{\M}/\hbar$, and after simple algebraic
manipulations, we find
\begin{equation}
\label{sfin} \frac{E_{\M}t_{\New}}{\hbar}=\frac{
2\,(w/k_{\M})^{^{2}}\,[s(1)s(2)-s(0)s(3)] +
4\,(q_{\M}/k_{\M})[s^{\2}(1)-s(0)s(2)]}{
[s^{\2}(2)-s(0)s(4)]+4\,(q_{\M}/k_{\M})[s(1)s(2)-s(0)s(3)]
+4\,(q_{\M}/k_{\M})^{^2}[s^{\2}(1)-s(0)s(2)]}\,\,.
\end{equation}
For $q_{\M}\sim k_{\M}$, remembering that in the opaque limit
$s(n)\gg s(n+1)$,  we find
\[ \frac{E_{\M}t_{\New}}{\hbar}\approx \frac{
4\,(q_{\M}/k_{\M})[s(1)s(2)-s(0)s(3)]}{4\,(q_{\M}/k_{\M})^{^2}[s^{\2}(1)-s(0)s(2)]}=
\frac{q_{\M}}{k_{\M}},
\]
as anticipated by Eq.(\ref{s1}). In the limit $q_{\M}\ll k_{\M}$
the main contribution to the numerator and denominator comes from
the first term,
\[\frac{E_{\M}t_{\New}}{\hbar}\approx \frac{
2\,(w/k_{\M})^{^{2}}\,[s(1)s(2)-s(0)s(3)]}{
[s^{\2}(2)-s(0)s(4)]}\approx \frac{ 2\,[s(1)s(2)-s(0)s(3)]}{
[s^{\2}(2)-s(0)s(4)]}\,\,,
\]
reproducing Eq.(\ref{s2}).

\section*{\normalsize IV. NUMERICAL SIMULATIONS}

The {\em new} phase time formula (\ref{sfin}) has been tested for
an incoming gaussian wave packet,
\[ g(k) = \left\{ \begin{array}{cl}
\exp[\,-(k-k_{\0})^{^{2}}d^{^{2}}/\,4\,] & \,\,\,\mbox{for}\,\,\,0\leq k \leq k_{\M} \,\,,\\
0 & \,\,\,\mbox{otherwise}\,\,,
\end{array} \right.
\]
with a localization  $d=10/k_{\M}$, with a momentum distribution
 centered at $k_{\0}=k_{\M}/2$, and with an upper limit for the momentum
 distribution given by
 $k_{\M}=\mbox{KeV}\,/\,\hbar\, c$. The incoming electrons move
in the free region before the barrier with velocity
\[
v_{\0}= \hbar\,k_{\M}/2\,m=\sqrt{E_{\M}/2\,m}=10^{^{-3}}c\,\,.
\]
For a potential barrier of height $V_{\0}=E_{\M}(=1\,\mbox{eV})$,
the transit velocity is given, see  Eq.(\ref{vtra}), by
\[
v_{tra}=4.5\,\cdot\,10^{^{-3}}\,c\,\,.
\]
Numerical data are presented in Table 1. The time in which the
transmitted peak appears in the free region after the barrier is
calculated for different values of $L$. For increasing $L$, the
transit velocity, $v_{\tra}$, tends to a constant value which is
in excellent agreement with the analytic value obtained from
Eq.(\ref{vtra}).

To complete our numerical analysis, we have calculated the transit
velocity as a function of $k_{\M}L$ for different ratios of
$V_{\0}/E_{\M}$. The plots in Fig\,1 clearly show that such a
velocity tends to a constant value for $V_{\0}\to E_{\M}$. The
standard phase time, the new phase time and the numerical data are
plotted  in Fig.\,2. The new phase time  is in excellent agreement
with the numerical simulations. The standard phase time represents
a good approximation for increasing values of $V_{\0}/E_{\M}$.

\section*{\normalsize V. CONCLUSIONS}

The growing interest in understanding tunneling times in quantum
physics stimulated the authors in looking for  a new analytic
formula of the phase time for wave packets transmission through
opaque barriers. After a brief review of the derivation of the
phase time formula, which is based on the stationary phase method,
we discuss some intriguing questions about its appropriate use. In
the opaque limit, the filter effect is responsible for a shift of
the mean value of the transmitted momentum. This allows to compute
directly the transmitted electronic density, and, consequently, by
taking the time derivative of this density, to find the time in
which the wave packet appears in the free region after the barrier
potential. The {\em new} formula for the phase time  is in
excellent agreement with the numerical simulations and clearly
shows in which cases the standard phase time, calculated by the
stationary phase method, represents a good approximation for the
transit time. The most important goal of the paper is the proof
that, for wave packets whose upper limit of the momentum
distribution is very close to the barrier height, the phase time
is {\em proportional} to the barrier width.

Finally, we hope that this work will find readers not only among
the physicists interested in tunneling phenomena but also among
the specialists in related branches of natural sciences which use
the stationary phase method in their practical research.

\section*{\small \rm ACKNOWLEDGEMENTS}

The authors thanks the referees for their observations and Prof.
Pietro Rotelli for reading the revised version of the manuscript
and for his very useful suggestions. One of the authors (SdL) also
thanks the Department of Physics, University of Salento (Lecce,
Italy), for the hospitality and the FAPESP (Brazil) for financial
support by the Grant No. 10/02216-2.

\newpage

\begin{table}
\begin{center}
\begin{tabular}{|c||c||c|c|}\hline
$L$  & $t$ & $v_{tra}$ & {\sc Ana/Num}
\\ $[\,\hbar/\sqrt{2\,m\,V_{\0}}\,\,]$ & $[\,\hbar/\,V_{\0}\,]$ & $[\,\sqrt{V_{\0}/\,2\,m}\,\,]$& $\diamond$ \\
\hline \hline \,\,\,50  & 10.20 & 4.9013  & 91.81 $\%$
\\ \hline
100  & 21.41 & 4.6715  & 96.33 $\%$
\\ \hline
150  & 32.37 & 4.6338  & 97.11 $\%$
\\ \hline
200  & 43.28 & 4.6209  & 97.38 $\%$
\\ \hline
250  & 54.17 & 4.6150  & 97.51 $\%$
\\ \hline
300  & 65.05 & 4.6118  & 97.58 $\%$
\\ \hline
350  & 75.92 & 4.6099  & 97.62 $\%$
\\ \hline
400  & 86.79 & 4.6086  & 97.64 $\%$
\\ \hline
450  & 97.66 & 4.6078  & 97.66 $\%$
\\ \hline
500  & 108.53\,\,\, & 4.6072  & 97.67 $\%$
\\  \hline
\end{tabular}\\
\begin{flushright}
$\begin{array}{rcl}
\hbar/\sqrt{2\,m\,V_{\0}}& \approx & 2\,\mbox{\AA}\\
\hbar/\,V_{\0} & \approx & 0.66\,\cdot \, 10^{\mi \1 \5}\,\mbox{sec}\\
\sqrt{V_{\0}/\,2\,m} & \approx & 10^{^{-3}}c
 \end{array}$
\end{flushright}
\end{center}
\caption{Numerical data for a localized ($d=20\,\mbox{\AA}$) non
relativistic ($v_{\0}=10^{\mi \3}\,c$) electron with
$E_{\M}=V_{\0}$ which tunnels across a barrier potential,
$V_{\0}=1\,\mbox{eV}$. The time in which the transmitted peak
appears in the free region after the barrier (second column), is
calculated for different values of $L$. It is clear the linear
dependence on $L$ of the transmission time. Consequently,  the
barrier transit velocity (third column) defined by $L/t$, tends to
a {\em constant} value. The ratio between analytical and numerical
transit velocities is given in the last column, and shows, as
expected, an excellent agreement for increasing values of $L$.}
\end{table}

\newpage

\begin{figure}[hbp]
\hspace*{-2.5cm}
\includegraphics[width=19cm, height=22cm, angle=0]{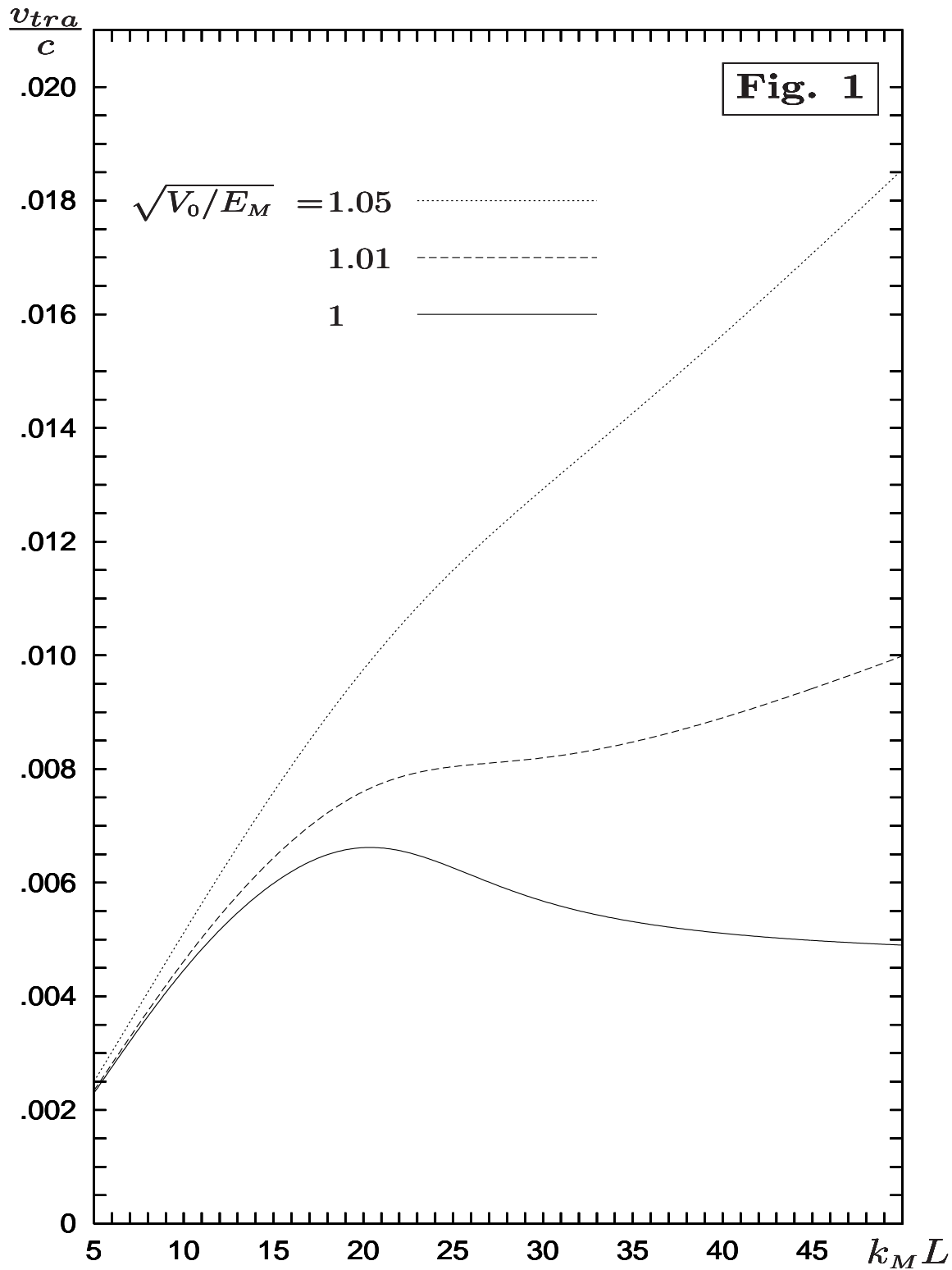}
\vspace*{-2cm}
 \caption{Transit velocities for an incoming (electron) wave packet centered in
 $k_{\0}=k_{\M}/2$ with localization determined by $k_{\M}d=10$
 as a function of $k_{\M}L$. For $V_{\0} \to E_{\M}$, the transit
 velocity tends to a constant value.}
\end{figure}

\newpage

\begin{figure}
\hspace*{-2.5cm}
\includegraphics[width=19cm, height=22cm, angle=0]{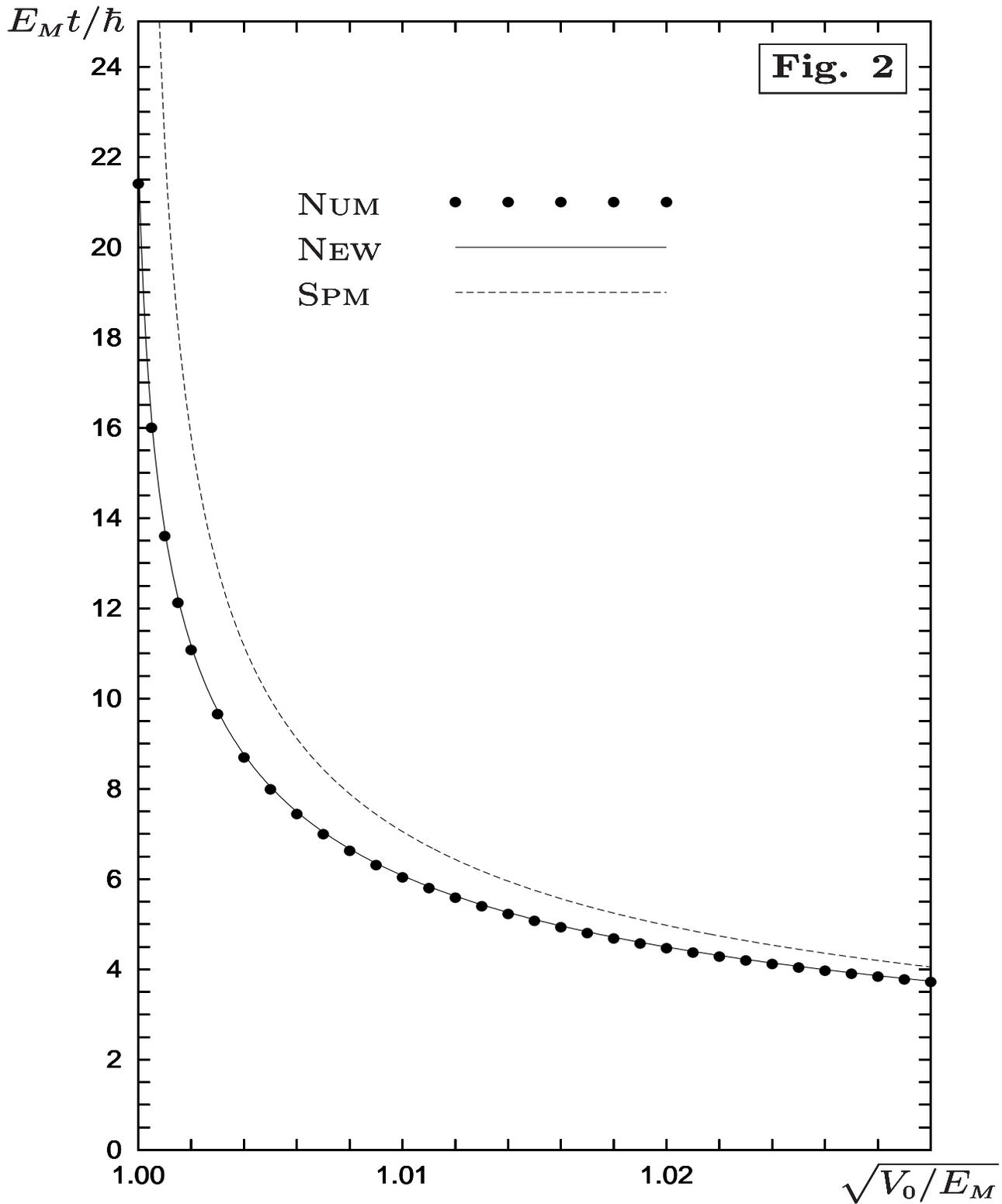}
\vspace*{-2cm}
 \caption{The new phase time formula ({\sc New}), the standard phase time formula obtained by
 using the stationary phase method ({\sc Spm}) and the numerical data ({\sc Num}) are plotted
 as  functions of $\sqrt{V_{\0}/E_{\M}}$ for a barrier width determined by $k_{\M}L=100$.
 The new phase time
formula is in excellent agreement with the numerical analysis. For
increasing $V_{\0}/E_{\M}$, the standard phase time formula
represents a good approximation. }
\end{figure}

\end{document}